\documentclass[prb,aps,twocolumn,showpacs,superscriptaddress,amsmath,amssymb]{revtex4}

\usepackage{graphicx}
\usepackage{psfrag}
\usepackage{epsfig}
\usepackage{color}
\usepackage{subfigure}
\definecolor{dred}{rgb}{0.7,0.0,0.0}
\usepackage{dcolumn}
\usepackage{bm}

\begin{document}

%
%

\title{On-site Atractive 
Multiorbital Hamiltonian for $d$-Wave Superconductors}

\author{Christopher B. Bishop}
 
\affiliation{Department of Physics and Astronomy,University of Tennessee,
Knoxville, TN 37966, USA} 
\affiliation{Materials Science and Technology Division,
Oak Ridge National Laboratory,Oak Ridge, TN 37831, USA}
 
\author{Guangkun Liu}
 
\affiliation{Department of Physics and Astronomy,University of Tennessee,
Knoxville, TN 37966, USA} 
\affiliation{Department of Physics, Beijing Normal University, Beijing 100875, China}
 
\author{Elbio Dagotto}
 
\affiliation{Department of Physics and Astronomy,University of Tennessee,
Knoxville, TN 37966, USA} 
\affiliation{Materials Science and Technology Division,
Oak Ridge National Laboratory,Oak Ridge, TN 37831, USA}

\author{Adriana Moreo}

\affiliation{Department of Physics and Astronomy,University of Tennessee,
Knoxville, TN 37966, USA} 
\affiliation{Materials Science and Technology Division,
Oak Ridge National Laboratory,Oak Ridge, TN 37831, USA}

\date{\today}

\begin{abstract}
{We introduce a two-orbital Hamiltonian on a square lattice that contains on-site attractive interactions 
involving the two $e_g$ orbitals. Via a canonical mean-field procedure similar to the one applied to 
the well-known negative-$U$ Hubbard model, it is shown that the new model develops $d$-wave ($B_{1g}$) superconductivity
with nodes along the diagonal directions of the square Brillouin zone.
This result is also supported by exact diagonalization of the model in a small cluster.
The expectation is that this relatively simple attractive model could be used 
to address the properties of multiorbital $d$-wave superconductors in the same manner that
the negative-$U$ Hubbard model is widely applied to the study of the properties of $s$-wave single-orbital superconductors.
In particular, we show that by splitting the $e_g$ orbitals and working at three-quarters filling, such that
the $x^2-y^2$ orbital dominates at the Fermi level but the $3z^2-r^2$ orbital contribution is nonzero, 
the $d$-wave pairing state found here phenomenologically reproduces several properties of the superconducting 
state of the high $T_c$ cuprates.
}
\end{abstract}
 
 
\maketitle
 
\section{Introduction}

Simple model Hamiltonians that can capture the basic aspects of the electronic collective states observed
in complex materials, such as in the cases of antiferromagnetism or superconductivity, 
are crucial to advance the theoretical understanding of these nontrivial phases and
to interpret and guide experimental efforts. The standard Hubbard and $t-J$ models have
successfully allowed for the study of the properties of antiferromagnetic compounds in the undoped limit~\cite{hub,tJ} while the negative-$U$
Hubbard model is a useful tool for the study of canonical $s$-wave
superconductors, from the BCS regime in weak coupling to the realm of Bose-Einstein condensation
in its strong coupling limit.\cite{uneg,uneg1,mohit} The discovery of $d$-wave superconductivity in the high $T_c$ cuprates
created the need for an equivalent simple Hamiltonian to analyze $d$-wave condensates.\cite{bednordz,dwave} 
While it is widely believed that upon doping both the Hubbard and $t-J$ models develop $d$-wave
superconductivity, this regime is difficult to study because the signals of superconductivity may be hidden by other
more dominant energy scales such as the superexchange $J$.
In fact, numerically the evidence
for long-range order superconductivity in these models is rather weak in realistic regimes of couplings. 
On the contrary, for the negative $U$ Hubbard model, even 
in small systems such as $2 \times 2$ lattices, the $s$-wave pairing tendencies 
are already clearly apparent.\cite{elbioreview,mynegU} 

For these reasons, several efforts have been devoted to develop the analogous of the negative-$U$ Hubbard 
model but for $d$-wave superconductors.\cite{ouruv,ourd,kivelson} The simplest approach is based on the single orbital case, 
to keep the number of degrees of freedom to a minimum. This rational is based on the fact that one single band, albeit 
composed of hybridized oxygen $p$ and copper $d$ orbitals, does define the Fermi surface of the high $T_c$ cuprates. 
However, in this case of a single orbital system, a pairing operator with $d$-wave symmetry has to locate the two electrons that
form the Cooper pair in two different lattice sites, as opposed to the negative-$U$ $s$-wave pairing operator that describes
a rotationally invariant pair of electrons with opposite spin on a single lattice site. 
While an attractive on-site potential readily allows the formation of on-site Cooper pairs in the negative-$U$ Hubbard
model, interactions that bond electrons in nearest neighbor sites, as required for $d$-wave pairing, may induce the formation of 
extended clusters of carriers that eventually can lead to phase separation rather than superconductivity, 
as argued in previous work.\cite{ouruv} It is believed that by fine tuning parameters, or including the effects of long-range Coulomb repulsion, eventually
pairing can be stabilized, but these interactions lead to models that are difficult to study. 
In addition, Hamiltonians where short range attraction
competes with long-range repulsion could also 
form complex structures such as stripes that may compete with pairing.\cite{emery,tranquada} 
While magnetism is considered a crucial 
factor in the mechanism that generates $d$-wave pairing in the cuprates,\cite{doug,bob,anderson} 
finding a simple effective model involving only charge and spin degrees of freedom that readily 
displays robust $d$-wave superconductivity remains ellusive.

The discovery of high $T_c$ superconductivity in the iron-based pnictides and selenides has provided a novel 
playground to investigate the potentially crucial role played by having many simultaneously active
degrees of freedom involved in the mechanism of superconductivity.\cite{Fe-SC,pengcheng}
While there are indications of either $s$- or $d$-wave symmetry in the superconducting order parameter of representative members
of this family of compounds, 
it is clear that the orbital degree of freedom must be included in the theoretical description. In fact, 
at least three $d$-orbitals contribute to determine the Fermi surfaces. When these orbital degrees of 
freedom are included, it is sometimes forgotten that they themselves contribute to determine the symmetry of the order parameters. In fact,  
the pairing operators for the pnictides are classified according to both their spatial and orbital symmetry properties.\cite{Wang,our2os,our2ol,our3o}

The key observation in this publication is that the addition of the orbital degree of freedom allows for the possibility 
of developing {\it on-site} pairing operators whose
symmetry is non trivial, namely non $s$-wave. 
More specifically, in this publication we will explicitly show that an on-site pairing operator with $d$-wave ($B_{1g}$) symmetry
can be constructed for a two-orbital system with hybridized bands on a square lattice. Moreover, we show that 
on-site inter-orbital pairing tendencies can be generated
via an on-site interorbital attraction and an effectively 
antiferromagnetic ``Hund coupling'' term. The  strength of the attraction and effective Hund coupling is tuned
with a single parameter $V$ which in turn determines the strength of the superconducting order. 
We believe that the weak coupling regime of this
model will allow for the study of the properties of generic $d$-wave superconductors 
in the same manner that the BCS limit can be studied with the
negative $U$ Hubbard model, while the strong coupling limit will unveil a novel unexplored regime where a Bose condensation of $d$-wave pairs dominates.
In other words, all the fruitfull investigations carried out in the past for the negative $U$ Hubbard model can be now revisited employing
a novel model with robust $d$-wave pairing, opening a broad avenue of research.
 
The paper is organized as follows: in Section II the Hamiltonian for the new multiorbital 
$d$-wave model is presented. A mean-field study of the Hamiltonian is performed
in Section III. Section IV is devoted to the exact diagonalization of the model in 
a small cluster and the calculation of the $d$-wave and $s$-wave pairing correlations.
A final discussion of the main results is presented in Section V.

\section{The MD Model}

The tight-binding term of the multiorbital $d$-wave model (dubbed the MD model) 
introduced here results from applying the Slater-Koster~\cite{SK} method to the
$x^2-y^2$ and $3z^2-r^2$ $d$-orbitals using a square lattice. It is well known that these are the two orbitals of relevance in the colossal
magnetoresistive manganites.\cite{oureview} 
Also several authors have considered these two same orbitals to model the cuprates: despite the fact that only one band of mostly $x^2-y^2$
character determines the Fermi surface, in practice this band is at least weakly 
hybridized with the $3z^2-r^2$ orbital.\cite{feiner,buda,saka,millis,jang} 
Using the Slater-Koster formalism we obtain
\begin{eqnarray}
H_{\rm TB}=-\sum_{{\bf i},\hat{\nu},\alpha,\alpha',\sigma}(t_{\alpha,\alpha'}c^{\dagger}_{{\bf i},\alpha,\sigma}c_{{\bf i}+\hat{\nu},\alpha',\sigma}+h.c.)
\nonumber\\
-{\mu}\sum_{{\bf i},\alpha}n_{{\bf i},\alpha}+\delta\sum_{{\bf i}}n_{{\bf i},2},
\label{htbrs}
\end{eqnarray}
\noindent where $c^{\dagger}_{{\bf i},\alpha,\sigma}$ creates an electron at site ${\bf i}$, orbital $\alpha$, and with spin projection $\sigma$. The orbital label
$\alpha=1$ (2) indicates the $x^2-y^2$ ($3z^2-r^2$) orbital. 
$\hat{\nu}$ is a unit vector that takes the values $\hat{\bf x}$ or 
$\hat{\bf y}$. The hoppings are given by $t_{1,1}=t_1$, $t_{2,2}=t_2$, and $t_{1,2}=\tau_{\nu}t_3$ 
where $\tau_{\hat{\bf x}}=1$ and $\tau_{\hat{\bf y}}=-1$ (note that this last sign difference is crucial to obtain the $d$-wave pairing).
While the actual amplitudes $t_i$ depend on the overlap of integrals, 
it is customary to consider them as free parameters that are chosen to reproduce the shape of the Fermi surface of the system to be studied.\cite{footman}
$\mu$ is the chemical potential and $n_{{\bf i},\alpha}$ is the number operator. The parameter $\delta$ breaks the energy degeneracy 
between the $e_g$ orbitals, as it occurs in the cuprates.
The tight-binding portion of the Hamiltonian can be written in momentum-space via the Fourier transform 
$c^{\dagger}_{{\bf j},\alpha,\sigma}={1\over{\sqrt{N}}}\sum_{\bf k}e^{-i{\bf k}.{\bf j}}c^{\dagger}_{{\bf k},\alpha,\sigma}$ becoming
\begin{eqnarray}
H_{\rm TB} =\sum_{{\bf k},\sigma}[T_{\bf k}^{11}c^{\dagger}_{{\bf k},1,\sigma}c_{{\bf k},1,\sigma}+T_{\bf k}^{22}c^{\dagger}_{{\bf k},2,\sigma}c_{{\bf k},2,\sigma}+
\nonumber\\
+(T_{\bf k}^{12}c^{\dagger}_{{\bf k},1,\sigma}c_{{\bf k},2,\sigma}+h.c.)],
\label{htbk}
\end{eqnarray}
\noindent with
\begin{equation}
T_{\bf k}^{11}=-2t_1(\cos k_x+\cos k_y)-\mu,
\label{e11}
\end{equation}
\begin{equation}
T_{\bf k}^{22}=-2t_2(\cos k_x+\cos k_y)+\delta-\mu,
\label{e22}
\end{equation}
\begin{equation}
T_{\bf k}^{12}=2t_3(\cos k_x-\cos k_y).
\label{e12}
\end{equation}

\begin{figure}[thbp]
\begin{center}
\includegraphics[width=8cm,clip,angle=0]{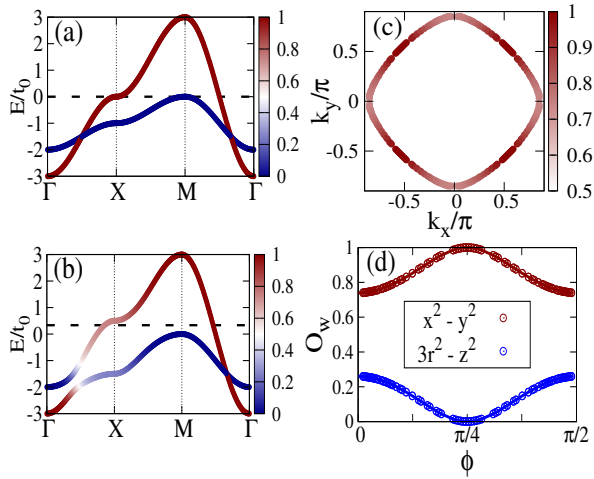}
\vskip 0.3cm
\caption{(color online) Band dispersion for the tight-binding term of the MD model discussed here: (a) is the non-hybridized case ($t_3=0$) and (b) the hybridized case
($t_3=-\sqrt{3}t_0/8$); the dashed line indicates the chemical potential position that leads to $\langle n\rangle=1.5$; 
(c) Fermi surface corresponding to the tight-binding Hamiltonian of panel (b). 
In panels (a), (b), and (c) the colors indicate the orbital composition of the bands, with
the scale ranging from 0 (blue, denoting $3z^2-r^2$) to 1 (red, denoting $x^2-y^2$). Panel (d) contains the orbital composition 
of the band that determines the Fermi surface of panel (c): the red (blue) circles indicate the weight of the $x^2-y^2$ ($3z^2-r^2$) orbital.}  
\label{fig1}
\end{center}
\end{figure}

While indeed the hoppings $t_i$ could be adjusted to reproduce the shape of a particular Fermi surface, 
to simplify the calculations we adopt the values
\begin{equation}
t_1={3t_0\over{4}},
\label{t1}
\end{equation}
\begin{equation}
t_2={t_0\over{4}},
\label{t2}
\end{equation}
\noindent and
\begin{equation}
t_3=-{\sqrt{3}t_0\over{8}},
\label{t3}
\end{equation}
\noindent so that all the hoppings $t_i$ can be expressed in terms of one single parameter
$t_0$ (again, note that the hopping $t_3$ also has a sign difference between the $x$ and $y$ directions, crucial
tor $d$-wave pairing). The tight-binding dispersion, with the energy in units of $t_0$, is shown in panel (a) of Fig.~\ref{fig1} for the 
non-hybridized special case where $t_3=0$. In this case, the band with the larger (smaller) bandwidth has pure $x^2-y^2$ ($3z^2-r^2$) character 
and it is indicated with a red (blue) line in the figure. 
The band dispersion for the hybridized case (nonzero $t_3$), that will be our main focus with regards to the existence of $d$-wave pairing, 
is shown in panel (b) of Fig.~\ref{fig1}. 
The colors indicate the mixture of the two orbitals in each of the bands. 
It can be observed that this orbital mixing is the strongest along the $\Gamma-X$ direction where a gap separates the bands 
that otherwise would cross as shown in the non-hybridized case. On the other hand,
along the diagonal of the Brillouin zone, $M-\Gamma$, there is no gap and the two bands still cross each other regardless of the value of $t_3$. 
The bandwidth is $W=6t_0$ as long as $|\delta|\le 3t_0$. Note that we have selected $\delta=-t_0$ and we 
have chosen a chemical potential $\mu$, indicated by a dashed line in the 
figures, that fixes the electronic density 
to three electrons per site, which means that the lower band is filled and the upper 
band is half-filled with an overall electronic density $\langle n \rangle =1.5$. 
It is clear that one band plays the dominant role to determine the Fermi surface shape shown in panel (c) for the hybridized case. 
The colors indicate that this Fermi surface is mostly of 
$d_{x^2-y^2}$ character, as in the case of the real cuprates, and the orbital mixing is maximized along the $\Gamma-X$ and $\Gamma-Y$ 
directions while it is zero along the diagonals. 

The Hamiltonian must transform as the $A_{1g}$ 
representation of the  $C_{4v}$ group of the square lattice. Then, since $\cos k_x-\cos k_y$ transforms like
$B_{1g}$ in Eq.~\ref{e12}, the product of operators $c^{\dagger}_{{\bf k},1,\sigma}c_{{\bf k},2,\sigma}$ also has to transform like $B_{1g}$. 
In fact, Eq.~\ref{htbk} can be rewritten as
\begin{equation}
H_{\rm TB}({\bf k})=\sum_{{\bf k},\sigma}\Psi^{\dagger}_{{\bf k},\sigma}\xi_{{\bf k}}\Psi_{{\bf k},\sigma},
\label{hcomp}
\end{equation}
\noindent where $\Psi^{\dagger}_{{\bf k},\sigma}=(c^{\dagger}_{{\bf k},1,\sigma},c^{\dagger}_{{\bf k},2,\sigma})$ and
\begin{equation}
\xi_{{\bf k}}=\epsilon_{\bf k}\sigma_0+\gamma_{\bf k}\sigma_1+\delta_{\bf k}\sigma_3,
\label{xi}
\end{equation}
\noindent with $\sigma_i$ the Pauli matrices and
\begin{equation}
\epsilon_{\bf k}={(T_{\bf k}^{11}+T_{\bf k}^{22})\over{2}}=-(t_1+t_2)(\cos k_x+\cos k_y)+{\delta\over{2}}-\mu,
\label{epsilonk}
\end{equation}
\begin{equation}
\delta_{\bf k}={(T_{\bf k}^{11}-T_{\bf k}^{22})\over{2}}=-(t_1-t_2)(\cos k_x+\cos k_y)-{\delta\over{2}},
\label{deltak}
\end{equation}
\noindent and
\begin{equation}
\gamma_{\bf k}=T_{\bf k}^{12}=2t_3(\cos k_x-\cos k_y).
\label{gammak}
\end{equation}

The expressions above establish that the orbital matrix $\sigma_1$ transforms 
like $B_{1g}$, while $\sigma_0$ and $\sigma_3$ transform like A$_{1g}$.

The orbital composition of the band that determines the Fermi surface is displayed in Fig.~\ref{fig1}~(d) as a function of 
the angle $\phi$, which is 0 when ${\bf k}_F$ is 
along the $x$ axis and $\pi/2$ when it is along the $y$ axis. The $B_{1g}$ character of the 
hybridization becomes clear since at $\phi=\pi/4$ the band is not hybridized, namely it consist of a pure $x^2-y^2$ orbital.
This means that a pairing operator of the form
\begin{equation}
\Delta^{(D)\dagger}_{\bf i}=c^{\dagger}_{{\bf i},1,\uparrow}c^{\dagger}_{{\bf i},2,\downarrow}-c^{\dagger}_{{\bf i},1,\downarrow}c^{\dagger}_{{\bf i},2,\uparrow},
\label{pairing}
\end{equation}
\noindent will transform as $B_{1g}$ and, therefore, it is a $d$-wave pairing operator. 

The next step is to construct an interaction term to be added 
to the Hamiltonian that would favor $d$-wave pairing. Based on the symmetry considerations above,
this term can be written as
\begin{equation}
H_{\rm int} = -V\sum_{\bf i}\Delta^{(D)\dagger}_{\bf i}\Delta_{\bf i}^{(D)},
\label{hint}
\end{equation}
\noindent by analogy with the attraction that leads to $s$-wave pairing 
($-U\sum_{\bf i}n_{{\bf i},\uparrow}n_{\bf{i},\downarrow}=-U\sum_{\bf i}\Delta^{(S)\dagger}_{\bf i}\Delta^{(S)}_{\bf i}$ 
with $\Delta^{(S)}_{\bf i}=c_{{\bf i},\downarrow} c_{{\bf i},\uparrow}$) in the negative-$U$
Hubbard model. Expanding Eq.~\ref{hint} in terms of the $c^{\dagger}_{{\bf i},\alpha,\sigma}$ operators it can be shown that
\begin{equation}
H_{\rm int} =2V\sum_{{\bf i}}{\bf S}_{{\bf i},1}\cdot{\bf S}_{{\bf i},2}-{V\over{2}}\sum_{{\bf i}}n_{{\bf i},1}n_{{\bf i},2},
\label{hJV}
\end{equation}
\noindent where ${\bf S}_{{\bf i},\alpha}={1\over{2}}\sum_{\sigma,\beta}c^{\dagger}_{{\bf i},\alpha,\gamma}\sigma^{\gamma,\beta}c_{{\bf i},\alpha,\beta}$ 
and $n_{{\bf i},\alpha}=\sum_{\sigma}c^{\dagger}_{{\bf i},\alpha,\sigma}c_{{\bf i},\alpha,\sigma}$.
Notice that these are precisely two of the terms that are already present in the interaction portion 
of the standard (repulsive) multiorbital Hubbard model, but now with couplings of opposite signs (qualitatively similar
as to how the sign of $U$ is reversed in the one-orbital Hubbard model to induce $s$-wave pairing).\cite{footmulti} 
Thus, an intuitive view of the interaction term introduced here is that it promotes spin singlet formation via an interorbital spin
antiferromagnetic coupling (i.e. the opposite of the canonical Hund's rule coupling that is ferromagnetic) 
as well as promoting pairing via an interorbital electronic attraction, the latter being similar to the intraorbital electronic 
attraction of the negative-$U$ Hubbard model. The total Hamiltonian for the multiorbital $d$-wave model is then
\begin{equation}
H_{\rm MD} = H_{\rm TB} + H_{\rm int}. 
\label{ham}
\end{equation}
As in the case of the negative-$U$ Hubbard model, the attractive interactions that appear in $H_{\rm int}$ should be understood as {\it effective} 
interactions that mimic the net effect of the (often complex) actual physical mechanism 
that causes the attraction in the $d$-wave channel. This real pairing mechanism may involve the spin, orbital, and/or lattice, and 
our model is an effective-model representation of those physical processes.

\section{Mean Field Study}

As in the case of the negative-$U$ Hubbard model, a simple mean-field approximation
is here expected to capture the essence of the ground state of the new proposed $d$-wave model. 
The interacting term $H_{\rm int}$ in momentum space is given by
\begin{equation}
H_{\rm int}=-{V\over{N}}\sum_{{\bf k,k'},\sigma,\sigma'}\sigma\sigma'c^{\dagger}_{{\bf k},1,\sigma}
 c^{\dagger}_{{\bf -k},2,-\sigma}c_{{\bf -k'},2,\sigma'}
c_{{\bf k'},1,-\sigma'}.
\label{attr}
\end{equation}
Introducing the standard mean-field
expectation values $b_{\bf k'}$=$
\langle c_{{\bf -k'},-\alpha,\downarrow}c_{{\bf k'},\alpha,\uparrow} \rangle$ 
and
$b^{\dagger}_{\bf k}=\langle c^{\dagger}_{{\bf k},\alpha,\uparrow}
 c^{\dagger}_{{\bf -k},-\alpha,\downarrow}\rangle$ and performing the substitution
$c^{\dagger}_{{\bf k},\alpha,\uparrow} c^{\dagger}_{{\bf -k},-\alpha,\downarrow}=
b^{\dagger}_{\bf k}+(c^{\dagger}_{{\bf k},\alpha,\uparrow}
c^{\dagger}_{{\bf -k},-\alpha,\downarrow}-b^{\dagger}_{\bf k})$
(and an analogous substitution for the product of annihilation
operators), the mean-field results are obtained. As usual, 
the fluctuations around the average given by
$(c^{\dagger}_{{\bf k},\alpha,\uparrow}
c^{\dagger}_{{\bf -k},-\alpha,\downarrow}-b^{\dagger}_{\bf k})$ are assumed to be small.
The resulting mean-field Hamiltonian is given by
\begin{equation}
H_{\rm MF}=\sum_{{\bf k}}\Psi^{\dagger}_{{\bf k}}H_{\bf k}\Psi_{{\bf k}}+\sum_{{\bf k}}(T_{\bf k}^{11}+T_{\bf k}^{22})+ 2V\Delta^2N,
\label{mf}
\end{equation}
\noindent where the generalized Nambu spinor is
$\Psi_{{\bf k}}=(c_{{\bf k},1,\uparrow},c_{{\bf k},2,\uparrow},c^{\dagger}_{{\bf-k},1,\downarrow},c^{\dagger}_{{\bf-k},2,\downarrow})^T$, 
while $H_{\bf k}$ is a $4 \times 4$ matrix given by 
\begin{equation}
H_{\bf k}=
\left(\begin{array}{cc}
\xi_{\bf k} & -V\Delta\sigma_1 \\
-V\Delta\sigma_1  & -\xi_{\bf k} 
\end{array} \right),
\label{6}
\end{equation}
\noindent where we have 
defined $\Delta=\langle c^{\dagger}_{{\bf k},\alpha,\uparrow}c^{\dagger}_{{\bf -k},-\alpha,\downarrow}\rangle=\langle c_{{\bf -k'},-\alpha,\downarrow}c_{{\bf k'},\alpha,\uparrow} \rangle$ and $N$ is
the number of sites.

\begin{figure}[thbp]
\begin{center}
\includegraphics[width=8cm,clip,angle=0]{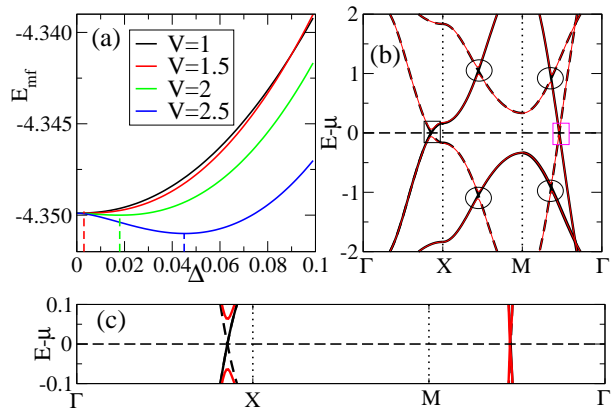}
\vskip 0.3cm
\caption{(color online) (a) Mean-field energy vs. the gap parameter $\Delta$ 
parametric with the strength of the attraction $V$, at $\langle n\rangle=1.5$. 
The value of $\Delta$ that minimizes the energy in each
case is indicated with the dashed lines; (b) Band structure along 
the directions $\Gamma-X-M-\Gamma$ using the mean-field approximation for the 
$H_{\rm MD}$ model for the cases $V=2$ (thin orange lines) and $V=0$ (black line), with 
the continuous lines denoting the tight binding dispersion and the dashed lines 
the replicas mean-field ``shadow'' bands. The ellipses indicate the gaps that open below and above the FS due to the 
orbital mixing, the black 
rectangle shows the gap at the FS, and  the magenta rectangle shows the $d$-wave node at the FS; 
(c) Detail corresponding to the black and magenta rectangles in panel (b).}
\label{mg}
\end{center}
\end{figure}

Similarly as for the case of magnetic order in the multiorbital Hubbard model,\cite{stephan} here we found that 
a finite value of the attractive coupling strength $V$ is needed to stabilize a nontrivial  solution with a
nonzero gap\cite{footnote} (this is different from the case of the negative-$U$ Hubbard model  
where a non-trivial solution occurs for any $|U|>0$ at any density). More specifically, we have found numerically
that a non-trivial solution appears for $V>t_0$ which is clearly still in the weak coupling regime since the bandwidth is $W=6t_0$. 
In panel (a) of Fig.~\ref{mg} the mean-field energy 
as a function of the gap parameter $\Delta$ is presented parametric with
$V$ at the electronic density $\langle n\rangle=1.5$. The particular value of $\Delta$ 
that provides the minimum mean-field energy is indicated for each value of the attraction.
The reason why pairing does not become stabilized 
with an infinitesimal attraction is due to the fact that 
if Eq.~\ref{6} is written in the base in which $\xi_{\bf k}$ is diagonal, the $2\times 2$ blocks $B=-V\Delta\sigma_1$ become
\begin{equation}
B=
\left(\begin{array}{cc}
V_{intra} & V_{inter} \\
V_{inter} & -V_{intra} 
\end{array} \right),
\label{7}
\end{equation}
\noindent where $V_{intra}=2V\Delta u_{\bf k}v_{\bf k}$ is the effective intraband pairing and  $V_{inter}=V\Delta (u^2_{\bf k}-v^2_{\bf k})$ is the effective
interband pairing and $u_{\bf k}$ and $v_{\bf k}$ are the elements of the unitarian matrix $U$ that performs the change of base transformation and satisfy
$u^2_{\bf k}+v^2_{\bf k}=1$.\cite{foot3}  The intraband potentials have opposite signs in each band. In addition, 
when $|k_x|=|k_y|$ the matrix $\xi_{\bf k}$ is diagonal, since the orbitals are not hybridized along this line, and
$u_{\bf k}=1$ while $v_{\bf k}=0$. In this case, the four eigenvalues of Eq.~\ref{6} at the FS are given by 
\begin{equation}
\lambda=\pm T_{\bf k}^{22}\pm\sqrt{(T_{\bf k}^{22})^2+V^2\Delta^2}.
\label{lambda}
\end{equation}
Since at the FS (see panel (d) of Fig.~\ref{fig1}) $u^2_{\bf k}-v^2_{\bf k}$ is always larger than 0, it is clear that $V_{inter}$ never vanishes preventing
the existence of a pure intraband attraction that would allow for the stabilization of a gap for any non-zero value of the attraction $V$.

As in the case of multiorbital magnetism, gaps in the band structure
appear not only at the Fermi surface (FS), but also at lower energies inside the Fermi sea. 
The mean-field band structure for the case of $V=2$ is indicated by the red lines in panel (b) of Fig.~\ref{mg} while the continuous (dashed) 
black lines denote the (``shadow'') band dispersion in the non-interacting case. It can be
observed that the interorbital attraction opens a gap at the FS, as indicated by the black rectangle 
along the $\Gamma-X$ direction and shown in detail in panel (c) of the figure. However, a node 
remains along the diagonal direction $M-\Gamma$, as highlighted by the magenta rectangle in panel (b) and detailed in panel (c)
of the figure. Strictly speaking, Eq.~\ref{lambda} shows that at the node there is a small gap given by $2(T_{\bf k}^{22}-\sqrt{(T_{\bf k}^{22})^2+V^2\Delta^2})$ which
is negligible for small values of $V$, as in Fig.~\ref{mg}, but eventually the node will be removed in the strong coupling limit.\cite{foot5}

The internal gaps that appear both above 
and below the FS due to the interorbital interaction are indicated with ellipses. Notice that
the gap along the FS is modulated by a function $f(k_x,k_y)$ with nodes for $|k_x|=|k_y|$, which arises 
from the matrix elements of the change of base matrix that transforms the system from the orbital to the band representation.\cite{foot5}
\begin{figure}[thbp]
\begin{center}
\includegraphics[width=8cm,clip,angle=0]{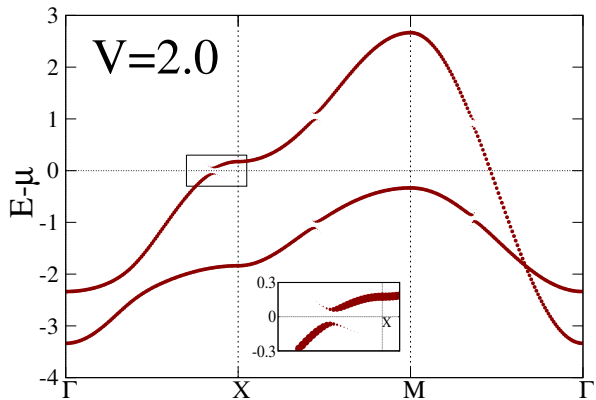}
\vskip 0.3cm
\caption{(color online) The spectral function A$({\bf k},\omega)$ along the directions $\Gamma-X-M-\Gamma$ in the mean-field approximation 
for the $H_{\rm MD}$ model, working at $V=2$. The inset highlights the weak intensity ``shadow'' spectral 
weight below and above the gaps opened by the attraction $V$. } 
\label{akw}
\end{center}
\end{figure}

We have also calculated the spectral function A$({\bf k},\omega)$ for the case of $V=2$. This mean-field spectral weight is shown in 
Fig.~\ref{akw}. It can be observed that at the locations of the gaps, namely at the FS but also below and above that FS, 
the spectral weight is reduced and ``shadow'' spectral
 weight appears across the gap. In other words, the single peak in the spectral function now splits into two. Notice that the opening of 
gaps located away from the FS is an effect caused by the interorbital interaction and it could explain the puzzling 
result recently observed in some iron superconductors where a superconducting gap appears in a band that is below the Fermi surface.\cite{gapbelow}

\section{Exact Diagonalization using 2$\times$2 clusters}

\begin{figure}[thbp]
\begin{center}
\includegraphics[width=8cm,clip,angle=0]{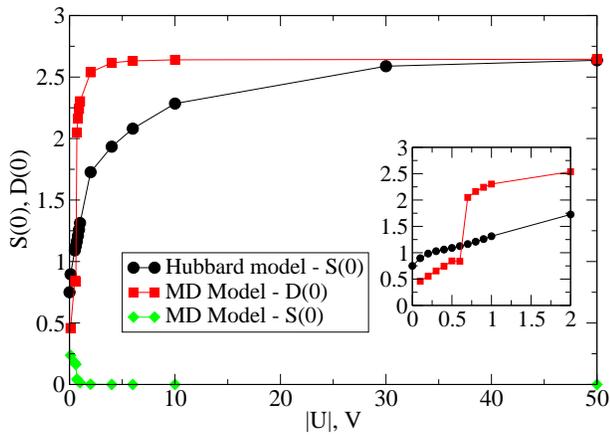}
\vskip 0.3cm
\caption{(color online) Exact diagonalization results using $2\times 2$ clusters. Shown is the Fourier transform at momentum zero, $D(0)$, of the $d$-wave pairing 
correlations for the MD model (with $\langle n \rangle =1.5$ i.e. $N$=12) vs. the on-site attraction strength $V$ (squares). 
The diamonds denote $S(0)$, the Fourier transform of the on-site $s$-wave pairing correlations. For comparison,
$S(0)$ for the negative $U$ Hubbard model is also shown varying the strength of the attraction $U$ (circles). 
The inset shows $D(0)$ for the MD model and $S(0)$ for the negative-$U$ Hubbard
model in the weak coupling regime. }
\label{fig4}
\end{center}
\end{figure}

It is well-known that the tendencies towards $s$-wave pairing are clear in the negative-$U$ Hubbard model even already in
a rather small $2\times 2$ cluster.\cite{mynegU} For this reason, we found  useful to perform an exact diagonalization of the novel 
MD model in this small cluster size (because the number of degrees of freedom now includes the orbital, this is the largest non-tilted
square cluster that can be fully diagonalized exactly). 
Working in subspaces with a fixed number of particles ranging from 0 to 16 we 
found the ground state energies and studied their behavior varying the chemical potential $\mu$ for several values of the attraction $V$. In 
Fig.~\ref{fig4} the squares indicate the zero-momentum Fourier transform of the $d$-wave pairing correlation 
functions $\langle \Delta^{(D)}_{\bf i}\Delta^{(D)\dagger}_{{\bf i}+{\bf r}} \rangle$ 
for the case $N=12$, namely $\langle n \rangle = 1.5$ in the $2\times 2$ cluster as a function of the attraction $V$. For comparison, 
the circles indicate the results for the $s$-wave pairing operator 
in the negative-$U$ Hubbard model as a function of $|U|$ given by
the ${\bf k}=0$ Fourier transform of $\langle \Delta^{(S)}_U({\bf i})\Delta^{(S)\dagger}_U({\bf i}+{\bf r}) \rangle$ with $\Delta^{(S)\dagger}_U({\bf i})=c^{\dagger}_{{\bf i},\uparrow}c^{\dagger}_{{\bf i},\downarrow}$. 
In both models, the pairing operators expectation values rapidly increase with the attraction, 
but in the inset it can be noticed that in the negative-$U$ model the pairing 
monotonously increases with $|U|$ while in the MD model there is a jump at $V\approx 0.7$ with a monotonous increase only afterwards. 
This behavior is in qualitative agreement with the mean-field result indicating 
that the superconducting state is stabilized only at a finite value of $V$ of order unity.
For comparison the onsite intraorbital $s$-wave pairing correlations for the MD model using the pairing operator 
$\Delta^{(S)\dagger}_{\bf i}=\sum_{\alpha}c^{\dagger}_{{\bf i},\alpha,\uparrow}c^{\dagger}_{{\bf i},\alpha,\downarrow}$ were also calculated 
(see diamonds in the figure). Clearly, there is no pairing in the $s$-wave channel, as expected. 

\begin{figure}[thbp]
\begin{center}
\includegraphics[width=8cm,clip,angle=0]{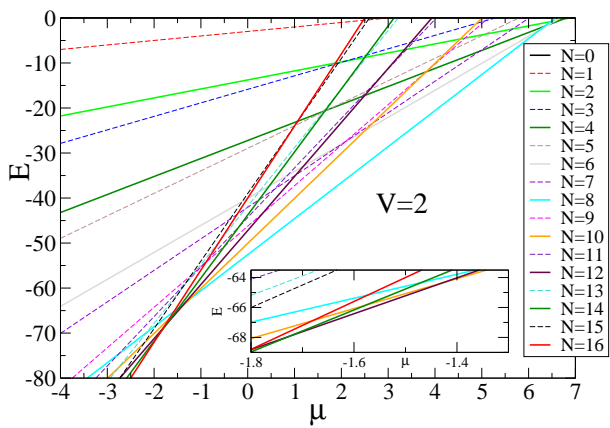}
\vskip 0.3cm
\caption{(color online) Exact diagonalization results corresponding to a $2\times 2$ cluster. Shown are the lowest energy levels for each possible value of $N$ 
as a function of the chemical potential for the MD model. The inset allows to see that
the states with $N$ = 14, 12, and 10 can be stabilized as ground states by the chemical potential for $V=2$.}
\label{fig5}
\end{center}
\end{figure}

Additional evidence of pairing in the MD model is obtained by studying the behavior of the ground state 
energy varying the chemical potential. In
Fig.~\ref{fig5} the ground state energies for the states with even (odd) number of particles are indicated with a straight (dashed) line. As in the 
negative-$U$ Hubbard model, only states with {\it even} number of particles are stable, 
indicating that the system displays pairing tendencies at all densities. The inset shows in more detail that 
all states with even number of particles can be stabilized with an adequated chemical potential tuning 
suggesting that the system does not have phase separation, a problem previously observed in proposed 
$d$-wave models involving nearest-neighbor attraction, as opposed to the on-site attractions used here. 
We have also verified explicitly, by inspection of the wave functions, that the relative symmetry between all the $N$-even 
ground states is $B_{1g}$,\cite{foot2} as expected.

\section{Discussion}

In this publication, we have presented a two-orbital Hamiltonian $H_{\rm MD}$ that can generate on-site $d$-wave superconductivity due to the non-trivial symmetry of the 
overlap integrals between hybridized orbitals that form the bands at the Fermi surface. In particular, this minimum model for $d$-wave pairing 
contains two $e_g$ orbitals on a square lattice. 
Via a canonical mean-field calculation, we have shown that this model, that contains an attractive 
on-site interorbital interaction and 
an effective antiferromagnetic Hund interaction, indeed supports $d$-wave superconductivity if the orbitals $x^2-y^2$ and $3z^2-r^2$ are hybridized and  non-degenerate. 
In multiorbital materials, 
it is possible that interorbital pairing occurs at the Fermi surface. Moreover, 
it was shown that the interorbital attraction also opens gaps away from the Fermi surface, a phenomenon already experimentally observed, but not yet 
explained, in the pnictides.\cite{gapbelow} In addition, in analogy with the well known negative-$U$ Hubbard model, 
that despite the local attraction can be used to 
study phenomenologically BCS superconductors with extended pairs in real space, 
it is expected that this new simple model could be applied to the phenomenological study 
of the properties of $d$-wave superconductors because the on-site character of the interactions 
in the ${\rm MD}$ model readily stabilizes the $d$-wave superconducting 
state. This is to be contrasted with more physically realistic, but far more challenging, models in which $d$-wave pairing is expected 
to result from a fine tuning of the competition between long-range Coulomb repulsion
and a short-range attraction induced by antiferromagnetism. In this context
complex extended structures, such as stripes or inhomogeneous states, can be formed as observed both in the cuprates and in the 
colossal magnetoresistive manganites,\cite{emery,oureview} and they tend to compete with uniform superconductivity.

It is important to remark that the symmetry of the pairing order parameter in the ${\rm MD}$ model can be changed by modifying the lattice geometry or the
orbitals involved. For example, if the $d_{xz}$ and $d_{yz}$ orbitals are considered, still using a square lattice, the symmetry of the on-site order
parameter becomes $B_{2g}$, i.e., with nodes along the $x$ and $y$ axes of the Brilloun zone.\cite{our2os,our2ol} 
Also note that while we have focused on  electronic density $\langle n\rangle=1.5$ in order to ensure a single band FS, the
$d$-wave state is stabilized for all other densities as it is the case in the negative-$U$ Hubbard model. 
The addition of hoppings beyond nearest-neighbor sites 
to the tight-binding portion of the Hamiltonian can be used to fine-tune the shape
of any desired Fermi surface, as long as the hoppings are compatible with the constraints
imposed by the Slater-Koster analysis.

\section{Acknowledgments}

The authors gratefully acknowledge discussions with Cristian Batista.
This work was supported by the National Science Foundation, under Grant No. DMR-1404375.


\begin{thebibliography}{10}
\bibitem{footmulti} Notice that in the multiorbital Hubbard Hamiltonian the relationship $U=U'+2J$ has to be satisfied to maintain 
the invariance under rotations for degenerate orbitals. However, in the model discussed here the two orbitals are not degenerate.
\bibitem{foot5} We observe that the node along the diagonal is lifted for values of $V\approx 4$ when the value of the mean-field chemical potential 
becomes different from the non-interacting one.
\bibitem{foot2} We found that the ground state with $N$ particles has $A_{1g}$ ($B_{1g}$) symmetry if $N/2$ is even (odd). This means that the states with $N$ and $N\pm 2$ particles 
have to be connected by a pairing operator with $B_{1g}$ symmetry.
\end{thebibliography}
\end{document}